\documentclass[12pt]{article}
\usepackage[top = 2 cm, bottom = 2.8 cm, left = 2.5 cm, right = 1.8 cm]{geometry}
\usepackage{authblk, cite, color, amssymb, amsmath}
\usepackage[colorlinks = true, linkcolor = blue, citecolor = red]{hyperref}
\usepackage[dvipsnames]{xcolor}

\begin{document}

\title{Conformal (2+4)-Braneworld}

\author{\bf Merab Gogberashvili}
\affil{\small Javakhishvili Tbilisi State University, 3 Chavchavadze Avenue, Tbilisi 0179, Georgia \authorcr
Andronikashvili Institute of Physics, 6 Tamarashvili Street, Tbilisi 0177, Georgia}

\maketitle

\begin{abstract}

It is considered the 6D brane model where matter is trapped on the surface of a (2+4)-hyperboloid, as is suggested by the geometrical structure behind the 4D conformal group. The effective dimension of the bulk space-time for matter fields is five, with the extra space-like and time-like domains. Using the embedding theory the presence of the familiar factorizable 5D brane metrics in the both domains is shown. These metrics with exponential warp factors are able to provide with the additional reduction of the effective space-time dimensions down to four. It is demonstrated that the extra (1+1)-space is not simply connected and there is a gap in the range of the extra coordinates. This can explain stability of the model in the domain with the time-like effective fifth dimension and the appearance of the cosmological constant due to the tachyon condensation. It is found that the model exhibits orbifold symmetry and thus is free from the fermion chirality problem.

\end{abstract}
\vskip 5mm


There has been great interest recently in braneworld scenario \cite{brane-1, brane-2} with various space-like and time-like extra dimensions (see reviews \cite{Reviews-1, Reviews-2, Reviews-3, Reviews-4}). In this paper we want to consider the specific 6D model where matter fields are localized on the surface of a (2+4)-hyperboloid, the construction which is dictated by the geometry of the conformal group in ordinary Minkowski space-time \cite{Dir, PR}. 6D models with this signature were investigated in \cite{PI-1, PI-2} for the case of compact extra dimensions and in \cite{GM-1, GM-2} in the context of brane models.

Unlike the previous (2+4)-models \cite{GM-1, GM-2}, where matter fields are directly localized on the brane, here we consider the intermediate trapping of matter on the specific one-sheet 5D-surface and show that the induced metric in this case contains exponential warp factors, which provide additional reduction to the effective 4D space-time.

It is known that the non-linear 15-parameter conformal group, the symmetry group of main massless equations of physics, can be written as linear Lorentz-type transformations with the invariant (2+4)-form \cite{Dir, PR}:
\begin{equation} \label{6D-cone}
\eta_{\mu\nu}Y^\mu Y^\nu + T^2 - X^2 = 0 ~, ~~~~~~~ (\mu, \nu = 0,1,2,3)
\end{equation}
where $\eta_{\mu\nu} = {\rm diag}(1,-1,-1,-1)$ denotes the 4D Minkowski metric tensor. Here $T$ and $X$ are extra time-like and space-like coordinates and the 4D Minkowski coordinates, $x^\nu$, correspond to 6D ones as,
\begin{equation} \label{x-Y}
x^\nu = \frac \epsilon 2 \frac {Y^\nu}{(T - X)} ~, ~~~{\rm or}~~~ x^\nu = \frac \epsilon 2 \frac {Y^\nu}{(T + X)} ~,
\end{equation}
where the parameter $\epsilon$ is the radius of the 6D hyperboloid (\ref{6D-cone}). Using the expressions (\ref{x-Y}) one can bring the equation of the null 6D conformal hyperboloid (\ref{6D-cone}) to the forms:
\begin{equation} \label{6D-hyper}
\eta_{\mu\nu}x^\mu x^\nu + \frac {\epsilon^2}{4}\frac {(T + X)}{(T - X)} = 0 ~, ~~~{\rm or}~~~ \eta_{\mu\nu}x^\mu x^\nu + \frac {\epsilon^2}{4}\frac {(T - X)}{(T + X)} = 0 ~.
\end{equation}

One of the transformations from the conformal group is the inversions,
\begin{equation} \label{conformal}
x'_\mu = \varepsilon \frac {x_\mu}{\eta_{\mu\nu}x^\mu x^\nu} ~,
\end{equation}
where $\varepsilon$ is some constant with the dimension of length$^2$. This transformation leaves the 6D form (\ref{6D-cone}) invariant if we will set $\varepsilon = \epsilon^2/4$ and will conduct the following transformations of the 6D coordinates:
\begin{equation} \label{inversion}
Y'_\mu = Y_\mu~, ~~~~~
\begin{cases}
T' = - T ~, \\
X' = - X ~.
\end{cases}
\end{equation}
Thus to the 4D inversions (\ref{conformal}), the two nonequivalent discrete transformations correspond with separate reflections of the temporal and spatial extra coordinates. The inversions (\ref{conformal}) with different signs,
\begin{equation} \label{epsilon}
\varepsilon = \pm \frac {\epsilon^2}{4}~,
\end{equation}
transform into each other the two non-equivalent representations of the 4D coordinates (\ref{x-Y}).

The conformal surface (\ref{6D-cone}) can be represented by some solution of the 6D Einstein equations with the delta-like source, $\sim \delta \left(\eta_{\mu\nu}Y^\mu Y^\nu + T^2 - X^2\right)$, if we identify the conformal parameter $\varepsilon$ with the radius of the hyperboloid $\epsilon$, as in (\ref{epsilon}). In this picture for the matter fields localized on (\ref{6D-cone}), the effective dimension of the bulk space-time will be five.

The space-time on the one sheet hyperboloid (\ref{6D-cone}) is not flat and has complex structure. One useful method to show this is the embedding theory \cite{eisenhart}. Depended on the sign of $T^2 - X^2$, the surface (\ref{6D-cone}) has two different embedment in a 7D space of the constant radius $\epsilon$,
\begin{eqnarray}
&\eta_{\mu\nu}Y^\mu Y^\nu - |T^2 - X^2| = -\epsilon^2~, ~~~~~~ (T^2 - X^2 < 0) \label{dS}\\
&\eta_{\mu\nu}Y^\mu Y^\nu + |T^2 - X^2| = \epsilon^2~. ~~~~~~~ (T^2 - X^2 > 0) \label{AdS}
\end{eqnarray}
Therefore the ($T\pm X$)-plains on the surface (\ref{6D-cone}) separate the domains with the positive and negative parameters, $\pm \epsilon^2$. Inside the causal cones of these plains we have the dS space (\ref{dS}), while outside the AdS space (\ref{AdS}), two simply-connected hyper-spaces of constant curvature. In the embeddings (\ref{dS}) and (\ref{AdS}) we cannot change the signs at the right sides, since the replacement of $\epsilon^2$ with $-\epsilon^2$ leads to the hyperboloid of two sheets, i.e. the conformal surface (\ref{6D-cone}) became not simply connected in this case. For the fields localized on the 6D hyperboloid (\ref{6D-cone}), the inversions with different sign (\ref{epsilon}) correspond to the transformations of (\ref{dS}) and (\ref{AdS}) spaces into each other.

To find the metrics of the 5D conformal surfaces (\ref{6D-hyper}), in the domains corresponding to the positive and negative inversion parameter $\varepsilon$, let us introduce the two new variables,
\begin{equation} \label{xi}
\xi_\pm = - \epsilon \ln \left( \frac 2\epsilon |T \pm X| \right) ~.
\end{equation}
Using embedding theory \cite{eisenhart}, it can be shown that the line element of the 6D hyperboloid (\ref{6D-cone}), in the domains (\ref{dS}) and (\ref{AdS}) can be brought to the two different kinds of 5D brane metrics of the models \cite{brane-1, brane-2}, and thus is able to reduce effective number of dimensions down to four.
\begin{itemize}
\item{In the dS domain (\ref{dS}) we have,
\begin{equation} \label{space-like}
ds^2 = \eta_{\mu\nu}dY^\mu dY^\nu + dT^2 - dX^2 = e^{-2\xi_+/\epsilon}\eta_{\mu\nu} dx^\mu dx^\nu - d\xi_+^2 ~.
\end{equation}
This embedding is done by the functions \cite{Embed}:
\begin{eqnarray} \label{Space-Em}
Y_\alpha &=& e^{-\xi_+/\epsilon} x_\alpha ~, \nonumber \\
T &=& \left( \frac{\epsilon}{4} - \frac 1\epsilon \eta_{\mu\nu}x^\mu x^\nu \right) e^{-\xi_+/\epsilon} + \epsilon e^{\xi_+/\epsilon} ~, \\
X &=& \left( \frac{\epsilon}{4} + \frac 1\epsilon \eta_{\mu\nu}x^\mu x^\nu  \right) e^{-\xi_+/\epsilon} - \epsilon e^{\xi_+/\epsilon} ~. \nonumber
\end{eqnarray}
}
\item{For the case of AdS space (\ref{AdS}) the reduction has the form:
\begin{equation} \label{time-like}
ds^2 =\eta_{\mu\nu}dY^\mu dY^\nu + dT^2 - dX^2 = e^{-2\xi_-/\epsilon}\eta_{\mu\nu} dx^\mu dx^\nu + d\xi_-^2 ~,
\end{equation}
which is done by the functions:
\begin{eqnarray} \label{Time-Em}
Y_\alpha &=& e^{-\xi_-/\epsilon} x_\alpha ~, \nonumber \\
T &=& \left( \frac{\epsilon}{4} - \frac 1\epsilon \eta_{\mu\nu}x^\mu x^\nu  \right) e^{-\xi_-/\epsilon} - \epsilon e^{\xi_-/\epsilon} ~, \\
X &=& \left( \frac{\epsilon}{4} + \frac 1\epsilon \eta_{\mu\nu}x^\mu x^\nu  \right) e^{-\xi_-/\epsilon} + \epsilon e^{\xi_-/\epsilon} ~. \nonumber
\end{eqnarray}
}
\end{itemize}
Using the constraints (\ref{dS}) and (\ref{AdS}) one can find that the inverse expressions of 5D coordinates ($x^\nu$ and $\xi_\pm$) by the 6D embedding functions ($Y^\nu$, $T$ and $X$) for the both cases are done by the expressions (\ref{x-Y}) and (\ref{xi}).

The 5D metrics in (\ref{space-like}) and (\ref{time-like}) (with the space-like and time-like fifth coordinates) represent the brane solutions of the 6D Einstein equations with the positive and negative cosmological constants (the dS and AdS domains of the (1+1)-plain spanned by $T$ and $X$). These 5D spaces with time-like and space-like fifth components, $\xi_\pm$, correspond to the invariant forms of the $O(2,3)$ and $O(1,4)$ rotational subgroups of the conformal group $O(2,4)$, respectively.

Now let us emphasize that the single 6D domain, (\ref{dS}) or (\ref{AdS}), is not enough for the reproduction of the whole values of the 4D interval,
\begin{equation}
-\infty < \eta_{\mu\nu} Y^\mu Y^\nu < \infty ~.
\end{equation}
Thus to describe conformal invariant 4D fields we need to use the both time-like and space-like domains of the extra ($T-X$)-plain, which are connected by the inversions (\ref{conformal}). The spaces (\ref{dS}) and (\ref{AdS}) also are connected through the reflections,
\begin{equation}
\eta_{\mu\nu} Y^\mu Y^\nu \longleftrightarrow - \eta_{\mu\nu} Y^\mu Y^\nu ~,
\end{equation}
since the 6D surface (\ref{6D-cone}) is symmetric under this transformation. This means that the extra dimensional spaces in (\ref{dS}) and (\ref{AdS}) are not simply connected:
\begin{itemize}
\item{For (\ref{dS}) the time-like and space-like 4D intervals are in the domains:
\begin{eqnarray}
- \epsilon^2 \leq &\eta_{\mu\nu} Y^\mu Y^\nu& \leq 0 ~, ~~~~~ (0 \leq |T^2-X^2| \leq \epsilon^2) \nonumber \\
\epsilon^2 \leq &\eta_{\mu\nu} Y^\mu Y^\nu& < \infty  ~. ~~~~~  (|T^2-X^2| \geq 2\epsilon^2)
\end{eqnarray}
}
\item{For (\ref{AdS}) the 4D intervals are distributed between the domains:
\begin{eqnarray}
0 \leq &\eta_{\mu\nu} Y^\mu Y^\nu& \leq \epsilon^2 ~,   ~~~~~ (0 \leq |T^2-X^2| \leq \epsilon^2) \nonumber \\
-\infty < &\eta_{\mu\nu} Y^\mu Y^\nu& \leq - \epsilon^2  ~. ~~~~~  (|T^2-X^2| \geq 2\epsilon^2)
\end{eqnarray}
}
\end{itemize}
We can conclude that the range:
\begin{equation} \label{gap}
\epsilon^2 \leq |T^2-X^2| \leq 2\epsilon^2
\end{equation}
is excluded for the 4D fields localized on the 6D hyperboloid (\ref{6D-cone}). So there is a gap in the values of $T^2 - X^2$, and thus of $\xi_\pm$, which covers only the part of the surface (\ref{6D-cone}), as in the models \cite{Kady-1, Kady-2, Kady-3, DeSitt-1, DeSitt-2, DeSitt-3}.

The fact that some part of the bulk space is unreachable for matter fields may be useful to solve the vacuum decay problem in the domain (\ref{time-like}) with the effective time-like extra dimension and at the same time is able to explain the appearance of the cosmological constant on the brane. It is known that theories with extra time-like dimensions have several pathological features, like the existence of excitations with negative norms \cite{Yndu}. In our case, due to the gap (\ref{gap}) for the time-like extra coordinate $\xi_-$, a physical field lowers its energy only up to the characteristic for the 6D hyperboloid value and the model appears to be stable under uncontrolled tachyonic decay. This mechanism also can explain the appearance of the cosmological constant on the brane of the order of $\epsilon^{-2}$ by the tachyon condensation mechanism. Indeed, the tachyonic condensates of matter fields in allowed time-like zone $\sim \epsilon$, will create a positive cosmological constant and lead to the appearance of the effective AdS space on the brane, i.e. would not affect 4D dispersion relations of matter fields.

Now let us consider the problem of parity transformations of physical fields localized on the 6D hyperboloid (\ref{6D-cone}). In general, definition of parity transformations in spaces with even spatial dimensions is not obvious, since reflection of all of the coordinates is equivalent to a rotation. Usual solution is reflection of all coordinates but the last one \cite{Shimizu-1, Shimizu-2}. In the case of (2+4)-space the situation is even more complicate, number of the both, spatial and temporal coordinates are even. However, due to the localization on the surface (\ref{6D-cone}), for matter fields the bulk space-time is effectively 5-dimensional, with time-like and space-like extra dimensional domains. Then the 6D rotational symmetry is broken and the accompanied Lorentz transformations are absent and it becomes possible to define parity transformations by the reflections of all spatial and temporal coordinates.

It is known that for odd dimensions there exist two inequivalent representations of the Clifford algebra, depending on the choice of sign of the product of all independent gamma matrices. The charge-conjugated spinor field of one of the representations is related to the spinor field of second representation and vice versa \cite{Shimizu-1, Shimizu-2}. This leads to the doubling of spinor components in odd dimensions, since these two independent fields separately are not form the representations of the complete Lorentz group (including discrete transformations). In order to make an irreducible representation we need the both fields, on the contrary to the case of even dimensions where there is only one representation.

We notice that the conformal 6D hyperboloid (\ref{6D-hyper}) is invariant under combined reflections $T \to -T$ and $X \to - X$, while reflection of only one of the extra coordinates (\ref{inversion}) leads to the inversions (\ref{conformal}) with different sign of $\varepsilon$. So Lagrangians in our 6D model needs to be invariant with the orbifold symmetry,
\begin{equation} \label{orbifold}
L(x^\nu, T, X) = L(x^\nu, -T, -X)~,
\end{equation}
and should contain two types of fields, $\psi_+ (x^\nu, \xi_+)$ and $\psi_+ (x^\nu, \xi_-)$, which transform into each other by the reflections of $T$ or $X$. These fields are localized in the space-like and time-like domains of the extra (1+1)-plain by the effective 5D brane solutions (\ref{space-like}) and (\ref{time-like}).

Now consider the fermion chirality problem in 5D \cite{5D-Chir-1, 5D-Chir-2,  5D-Chir-3}, which is that there is no a 5D parity operator, and hence 5D spinors are Dirac spinors. If every 4D Weyl spinor must come from a independent 5D Dirac spinor, then we end up doubling the number of degrees of freedom of our 4D theory, because every left-handed 4D Weyl fermion would have a corresponding right-handed 4D Weyl fermion (obtained from the same 5D Dirac fermion) with exactly the same quantum numbers. If the 5D fermion transforms under some 5D gauge symmetry, then the left and right 4D Weyl modes transform identically under this gauge symmetry. Hence, such a scenario cannot correspond to the Standard Model, where the fermions are known to be chiral, under the gauge symmetry the left and right ones transform as doublets and singlets, respectively. A solution of the 5D chirality problem is to orbifold extra dimensions \cite{5D-Chir-1, 5D-Chir-2, 5D-Chir-3}. Under an orbifold projection any 5D fermionic field transforms even,
\begin{equation}
\psi_e (x^\mu, - \xi) \longleftrightarrow \psi_e (x^\nu, \xi)~,
\end{equation}
or odd,
\begin{equation}
\psi_o (x^\mu, - \xi) \longleftrightarrow - \psi_o (x^\nu, \xi)~.
\end{equation}
For the odd fermion fields
\begin{equation}
\psi_o (x^\mu, 0) = 0~,
\end{equation}
thus they do not have 4D zero modes. As we have shown above, the orbifold symmetry of our model (\ref{orbifold}) automatically follows from the structure of the conformal 6D hyperboloid (\ref{6D-hyper}). This symmetry kills off one of the chiral zero modes, since upon integrating out the $\xi$ coordinate the kinetic terms of fermions trivially have the usual Standard Model structure.


In summary, in this paper was considered the specific 6D braneworld, where matter is trapped on the surface of a (2+4)-hyperboloid. The setup of the model is dictated by the geometrical structure behind the 4D conformal group. For the fields which are localized on the (2+4)-surface the bulk space-time is effectively 5-dimensional, with the extra space-like and time-like domains. Using the embedding theory, the line elements in the both domains was found. These metrics appeared to represent the familiar 5D brane solutions with the exponential warp factors, which are able to reduce effective number of dimensions down to four. It is demonstrated that the extra (1+1)-space is not simply connected and there is a gap in the range of the extra coordinates. This can explain the stability of our construction, in spite of the introduction of the effective time-like dimension, and the appearance of the cosmological constant by the tachyon condensation mechanism. It was found also that the model exhibits orbifold symmetry and thus is free from the fermion chirality problem.


\end{document}